\documentclass[trackchanges, twocolumn, twocolappendix]{aastex701}

\begin{document}

\title{Metal-polluted Population III Galaxies and How to Find Them}

\author[orcid=0009-0006-4326-6097]{Elka Rusta}
\affiliation{Dipartimento di Fisica e Astronomia, Università degli Studi di Firenze, Largo E. Fermi 1, 50125, Firenze, Italy}
\affiliation{INAF/Osservatorio Astrofisico di Arcetri, Largo E. Fermi 5, 50125, Firenze, Italy}
\email[show]{elka.rusta@unifi.it}  

\author[orcid=0000-0001-7298-2478]{Stefania Salvadori}
\affiliation{Dipartimento di Fisica e Astronomia, Università degli Studi di Firenze, Largo E. Fermi 1, 50125, Firenze, Italy}
\affiliation{INAF/Osservatorio Astrofisico di Arcetri, Largo E. Fermi 5, 50125, Firenze, Italy}
\email{stefania.salvadori@unifi.it}  

\author[orcid=0000-0001-5487-0392]{Viola Gelli}
\affiliation{Cosmic Dawn Center (DAWN), Denmark}
\affiliation{Niels Bohr Institute, University of Copenhagen, Jagtvej 128, 2200 Copenhagen N, Denmark}
\email{viola.gelli@nbi.ku.dk}  

\author[orcid=0000-0001-7144-7182]{Daniel Schaerer}
\affiliation{Department of Astronomy, University of Geneva, Chemin Pegasi 51, 1290 Versoix, Switzerland}
\email{daniel.schaerer@unige.ch}  

\author[orcid=0000-0002-9889-4238]{Alessandro Marconi}
\affiliation{Dipartimento di Fisica e Astronomia, Università degli Studi di Firenze, Largo E. Fermi 1, 50125, Firenze, Italy}
\affiliation{INAF/Osservatorio Astrofisico di Arcetri, Largo E. Fermi 5, 50125, Firenze, Italy}
\email{alessandro.marconi@unifi.it}  

\author[orcid=0000-0002-3524-7172]{Ioanna Koutsouridou}
\affiliation{Dipartimento di Fisica e Astronomia, Università degli Studi di Firenze, Largo E. Fermi 1, 50125, Firenze, Italy}
\email{ioanna.koutsouridou@unifi.it}  

\author[orcid=0000-0002-6719-380X]{Stefano Carniani}
\affiliation{Scuola Normale Superiore, Piazza dei Cavalieri 7, 56126 Pisa, Italy}
\email{stefano.carniani@sns.it}

\correspondingauthor{Elka Rusta}

\begin{abstract}

Observing Population III (hereafter PopIII) galaxies, the hosts of first-generation stars, remains challenging even with the JWST. The current few candidates have been identified through the combination of a prominent HeII emission and the absence of metal lines, a well-known but extremely brief signature of metal-free systems.
Here, we accurately model the evolution of the emission from PopIII galaxies to increase the number of candidates in JWST observations.
To achieve this, we employ a locally calibrated galaxy-formation model that self-consistently follows the star formation and chemical evolution initiated by the first stars.
We find that PopIII galaxies can emit metal lines in their ``self-polluted'' phase, 
while galaxies host only metal-free stars but the gas has been chemically-enriched by the first supernovae. In this phase, PopIII galaxies have $\rm [OIII]/H\beta \approx 1$, which opens the pool of candidates to more easily detectable sources. We predict that the high HeII emission of PopIII galaxies can last up to $\rm \approx 20 \, Myrs$ and that it is partly maintained in the ``hybrid'' phase, when PopIII and PopII stars co-exist in the host galaxy. We propose novel diagnostics involving UV metal lines to select PopIII candidates in high-z JWST surveys. In JADES, we identify 9 candidate galaxies with $>25\%$ of their stellar mass in metal-free stars, showcasing the effectiveness of our method. Ultimately, the key to discovering PopIII galaxies could be to catch them during their first episodes of chemical enrichment.

\end{abstract}

\keywords{\uat{Population III stars}{1285} --- \uat{High-redshift galaxies}{734} --- \uat{Chemical enrichment}{225} --- \uat{James Webb Space Telescope}{2291}}

\section{Introduction} 
\label{sec:intro}

One of the most ambitious goals of JWST is detecting galaxies that host the first generation of metal-free stars, also known as \textit{Population III} (hereafter PopIII) galaxies. Despite our current ability to observe the first billion years of the Universe with unprecedented detail, there are only a handful of tentative PopIII candidates \citep{vanzella+23, maiolino+24, wang+24, fujimoto+25, Nakajima+25}. Observing PopIII galaxies is indeed extremely challenging. At high redshifts, $z \approx 30-15$, PopIII galaxies are expected to be the major contributors to the star formation (SF) rate density but to be tremendously faint because of their low stellar masses \citep[e.g,][]{bromm+11,Hartwig2022}. As $z$ decreases, PopIII galaxies become brighter but also increasingly rare, since they reside in pristine regions of the Universe. Thus, although PopIII galaxies might form down to $z \approx 3$ \citep[e.g.][]{pallottini2014simulating, Liu+20, Saccardi2023, zier+25}, to identify them in the plethora of normal (PopII) galaxies we need very clear, and long-lasting, spectral signatures.

The search for PopIII galaxies is currently based on the well-established spectral feature of metal-free systems: 
a strong HeII emission combined with the absence of metal lines \citep[e.g.][]{Schaerer2002, inoue+11}. While the latter is due to the putative pristine composition of the interstellar medium (ISM) surrounding PopIII stars, the strong HeII emission is directly linked to their metal-free nature. Pristine stars are hotter and denser than metal-enriched (PopII) stars, and thus they have harder spectra \citep[e.g.][]{ tumlinson+00, marigo+01}. 
Furthermore, due to the lack of heavy elements in their birth environment, PopIII stars are likely more massive than PopII stars \citep[e.g.][for a recent review]{tan+08, klessen+23}, with simulations predicting initial masses up to $\rm 10^3 M_\odot$ \citep[e.g.][]{hirano2014one, Susa2014} or even more extreme values of $\rm 10^5 M_\odot$ \citep[e.g.][]{hosokawa+13, nandal+05}. Stellar archeology studies confirm the massive nature of PopIII stars, suggesting a peak of their unknown Initial Mass Function (IMF) of a few tens of solar masses \citep[e.g.][]{deBen2017, Hartwig2018a,pagnini+23}. Since massive stars are intrinsically hotter, a top-heavy PopIII IMF implies an even harder spectrum. Thus, the large amount of energetic radiation produces the strong HeII recombination lines that ultimately constitute the clear spectral feature of PopIII galaxies. 

Still, only three PopIII candidates have been identified using this key spectral signature: a HeII-emitting clump with no associated metal lines at $z=10.6$ \citep{maiolino+24}, a strong HeII emitter at $z=8.16$ \citep[RX J2129–z8HeII,][]{wang+24}, and an extremely metal-poor complex 
at $z=6.6$ (LAP1, \citealt{vanzella+23}; LAP1-B, \citealt{Nakajima+25}).
How can we expand the sample of PopIII candidates? 

The main challenge in finding PopIII galaxies lies in the brief duration of their key spectral signatures. Indeed, the intensity of the HeII recombination lines rapidly decreases after $\approx3$~Myrs \citep[e.g.][]{Schaerer2002}. Moreover, the large masses of PopIII stars result in a relatively short lifespan, only $\approx (3-30)$~Myrs, after which they quickly pollute the surrounding environment with the metals forged in their brief lifetimes. 

PopIII stars ending their life as supernovae (SNe) imprint the surrounding gas with unique abundance patterns \citep[e.g.][]{Vanni+23}. These chemical signatures are used to study PopIII-enriched gas in distant absorbers \citep[e.g.][]{Maio2013, Vanni2024} but are neglected when studying the direct emission of PopIII galaxies, typically assumed to host metal-free gas \citep[e.g.][]{zackrisson+11, trussler+23, venditti+24, lecroq+25}. So far, we lack a comprehensive work that self-consistently combines spectral signatures of metal-free stars with diagnostics of PopIII-polluted gas. 

In this Letter, we investigate the evolution of galaxies hosting first-generation stars. To this end, we model their emission spectra throughout different evolutionary phases to understand how long we can still detect the spectral signatures of metal-free stars and address these questions: 
How does the emission of a PopIII galaxy evolve? Can we detect PopIII stars in a chemically-enriched galaxy? How do we identify ``hybrid'' galaxies that host both metal-free stars and the following metal-enriched generations?

\section{Methods}
\label{sec:methods}

We simulate PopIII galaxies with the locally calibrated galaxy-formation model {\tt NEFERTITI} (NEar FiEld cosmology: Re-Tracing Invisible TImes, \citealt{Koutsouridou2023}), summarized in Sec.~\ref{sec:2.1}. To produce their synthetic emission, we run photoionization models with the Cloudy code \citep[C23 release,][]{chatzikos+23,ferland+98}, as explained in Sec.~\ref{sec:2.2}.

\subsection{Modeling PopIII galaxies with NEFERTITI}
\label{sec:2.1}

\begin{figure*}[thbp]
    \includegraphics[width=0.53\hsize]{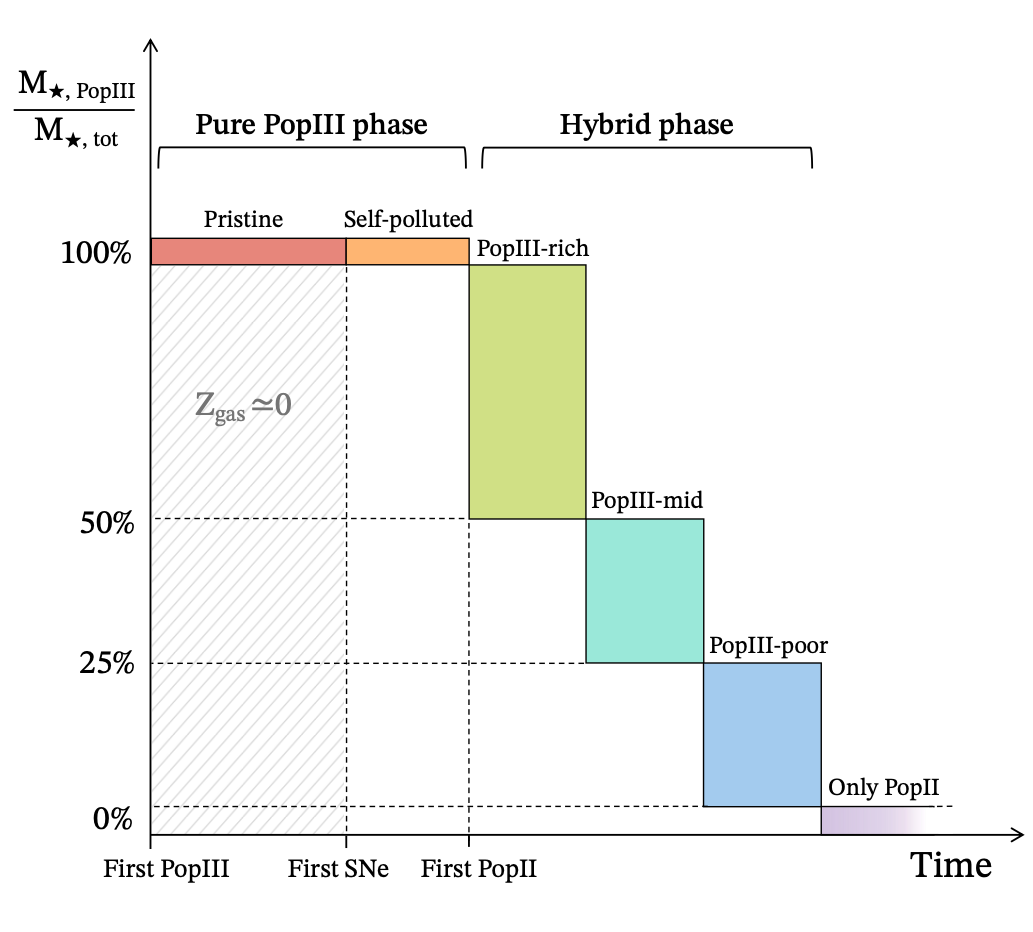} 
    \includegraphics[width=0.47\hsize]{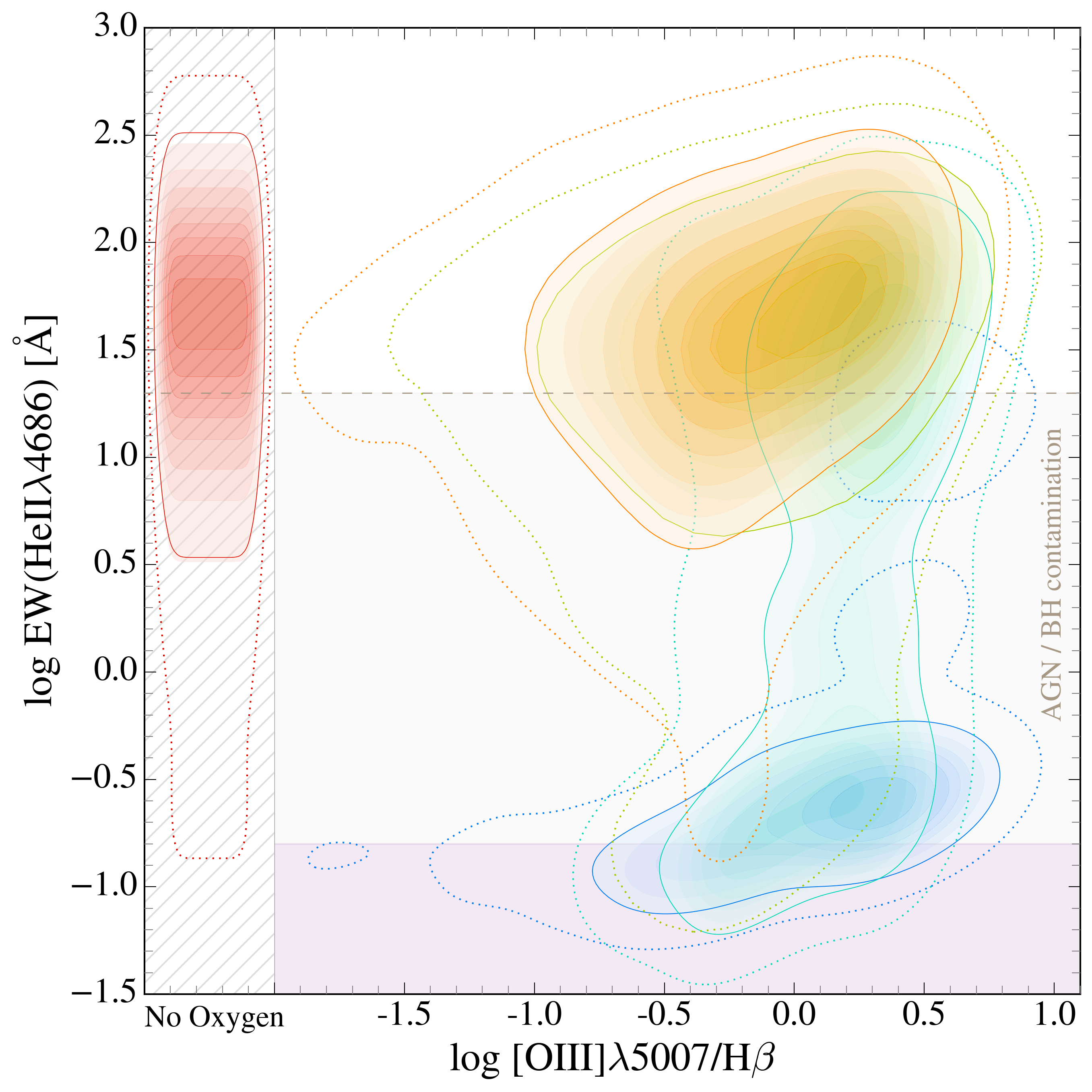} 
    \caption{\textit{Left:} Schematic representation of the evolutionary stages of a PopIII galaxy, based on the ratio between PopIII and total $\rm M_\star$. \textit {Right:} Density distributions of {\tt NEFERTITI} PopIII galaxies on the $\rm EW(HeII\lambda4686)$ versus $\rm [OIII]\lambda 5007/H\beta$ diagram 
    for $\rm logU=-1$. The colors represent different evolutionary stages, as shown in the left side scheme. The dotted (solid) lines include $95\%$ ($68\%$) of the population.
    The horizontal dashed line is the lower limit of the PopIII diagnostic from \citealt{Nakajima+22}, with the shaded area below highlighting the AGN and BH contamination zone. }
  \label{fig:fig1}
\end{figure*}

{\tt NEFERTITI} is a semi-analytical model for the formation and evolution of individual stars in early galaxies, specifically aimed at studying the properties of PopIII stars and their impact on galaxy evolution. 

Here we couple {\tt NEFERTITI} with a dark matter (DM) cosmological simulation of a Milky Way (MW) analog and calibrate it by reproducing a large set of present-day observations for the MW \citep{Koutsouridou2023}, including stellar archeology data such as the metallicity distribution function of Galactic halo stars \citep{Bonifacio2021}. 
This calibration is key because the most ancient and metal-poor stars 
likely formed in the gas enriched by the first PopIII SNe, allowing us to indirectly study PopIII galaxies.
For this reason, {\tt NEFERTITI} is a state-of-the-art model for understanding the evolution of PopIII galaxies.

{\tt NEFERTITI} grounds on previous semi-analytical models \citep{Salvadori2010, Salvadori2015} but is specifically designed to explore the unknown PopIII IMF and energy distribution of PopIII SNe \citep{Koutsouridou2023, Koustouridou2024}. Moreover, it employs a random sampling technique to handle weaker SF bursts that cannot fully populate the theoretical IMF in low-mass halos, which is particularly important for PopIII SF. 
We perform 5 runs of {\tt NEFERTITI} with stochastic PopIII IMF sampling and metal enrichment \citep{rossi2021ultra} to obtain a wide and diverse sample of PopIII galaxies.

At the start of the simulation, {\tt NEFERTITI} assumes that newly formed DM halos accrete zero-metallicity gas from the intergalactic medium (IGM), with a rate proportional to their DM growth. When the halo mass exceeds a minimum threshold \citep[as in][]{salvadori2009}, the first PopIII stars form. The evolution of each star is then tracked individually, following the proper timescale and chemical elements yielded\footnote{For PopIII stars we assume the yields from \citealt{Heger2002} and \citealt{Heger2010}, while for PopII stars those of \citealt{Limongi2018}.}, from C to Zn \citep{Koutsouridou2025}. 
Then, when the ISM metallicity exceeds the critical value, $\rm Z_{gas} > 10^{-4.5} \, Z_{\odot}$ \citep{deBen2017} normal PopII stars form.

After the SNe explode, part of the ISM metals and gas are injected into the IGM, where the total mass of metals increases over time and so $\rm Z_{IGM}={M_Z^{IGM}/M^{IGM}_{gas}}$. To account for the inhomogeneous mixing of metals into the IGM, we use the average evolution of the metal filling factor $\rm Q= V_{fill}/V_{tot}$ computed by \citealt{Salvadori2014} for MW-like galaxies. At each timestep, we randomly assign to a fraction $\rm Q$ of DM halos an enhanced IGM metallicity of $\rm Z_{IGM}/Q$, while the others still accrete pristine gas. With this method, PopIII SF can occur in pockets of pristine gas down to $z \approx 6$.

We adopt a Larson-type IMF, $\rm {\phi = m^{-2.35} e^{(-m_{ch}/M_{\odot})}}$ \citep{larson1998early}, for both PopIII and PopII/I stars. We assume that the mass of PopIII stars ranges between $\rm m_\star = [0.8,1000] \, M_{\odot}$ and has characteristic mass $\rm m_{ch}=10 \, M_{\odot}$, consistently with stellar archeology constraints \citep[see][]{Koutsouridou2023, Koustouridou2024}. For PopII/I stars we adopt $\rm m_\star = [0.1,100]\,  M_{\odot}$ and $\rm m_{ch} =0.35\, M_{\odot}$. 

In this Letter, we analyze the evolution from $z=17$ to $z=6$ of 1128 PopIII galaxies with total stellar masses $\rm M_\star = 10^{2-7} M_\odot$. At each timestep of 1 Myr, we have the following information from {\tt NEFERTITI}: total stellar/gas/DM mass, stellar population formed, total gas metallicity $\rm Z_{gas}$, gas chemical abundances, 
and masses of the individual stars formed in each PopIII SF episode.

\subsection{Modeling emission lines with CLOUDY}
\label{sec:2.2}

We build the stellar emission of PopIII galaxies according to {\tt NEFERTITI} predictions as follows. 
For each timestep, if the galaxy forms PopIII stars, we sum the emission of individual stars to account for the incomplete sampling of the IMF, using stellar spectra and evolutionary tracks 
from \citealt{Schaerer2002}. 
For PopII SF, we adopt an IMF-integrated spectrum, using \citealt{schaerer+03} for metal-poor stars ($\rm Z \approx 5 \cdot 10^{-4} \, Z_\odot$) and \citealt{zackrisson+11} for higher metallicities up to $\rm Z \approx Z_\odot$.
Then, we assemble the spectra of all the timesteps that compose the SF history of the PopIII galaxy up to the desired age \citep[as in][]{rusta+24}.

Then, we input these stellar spectra into the Cloudy photoionization code \citep{chatzikos+23} to get the total emission of each galaxy. For the gas nebula, we specify the chemical abundances of each element predicted by {\tt NEFERTITI}, which allows us to follow the metal enrichment of the medium self-consistently. We assume gas clouds with plane-parallel geometry, constant density, no dust, and a fiducial value for the neutral hydrogen density of $\rm 10^3 cm^{-3}$. For the ionization parameter U, we explore the following wide range of values: $\rm logU = [-3, -2, -1, -0.5, 0]$. The calculations are stopped when the electron fraction falls below 0.01. Ultimately, we have a set of $\approx 28\,000$ new photoionization models of PopIII galaxies.

\section{Results}
\label{sec:results}

We investigate how the emission of a PopIII galaxy evolves in response to the chemical enrichment of its gas and the onset of PopII SF. For this purpose, we explore emission line diagrams involving combinations of HeII and metal lines. 

\subsection{Evolution of a PopIII galaxy}
\label{sec:3.1}

We begin by identifying the key phases in the evolution of our simulated PopIII galaxies, schematized on the left side of Fig.~\ref{fig:fig1}. 
The first is the {\it pure PopIII} phase, characterized by a completely metal-free stellar population, and the second is the {\it hybrid phase}, in which PopIII stars coexist with PopII stars. These two main stages can be subdivided into more steps. Initially, the galaxy has pristine gas and forms PopIII stars for 3-12 Myr, depending on our stochastic IMF sampling. Then the gas is ``self-polluted'' by the first PopIII SNe, resulting in a galaxy with a {\it purely metal-free stellar population but chemically enriched gas}.
In the {\tt NEFERTITI }model, this phase lasts only for our time resolution of 1 Myr, because the Pop~III-enriched gas reaches $\rm Z_{cr}$ and immediately starts forming normal PopII stars. 
The hybrid galaxy then evolves from a {\it PopIII-rich} phase, when $\rm >50\% M_\star$ is metal-free, to an intermediate {\it PopIII-mid} phase. Then, after 2-5 Myrs, it enters into a {\it PopIII-poor} phase, when $\rm M_{\star, PopIII}<25\% M_\star$.

\begin{figure}[thpb]
    \includegraphics[width=1.\hsize]{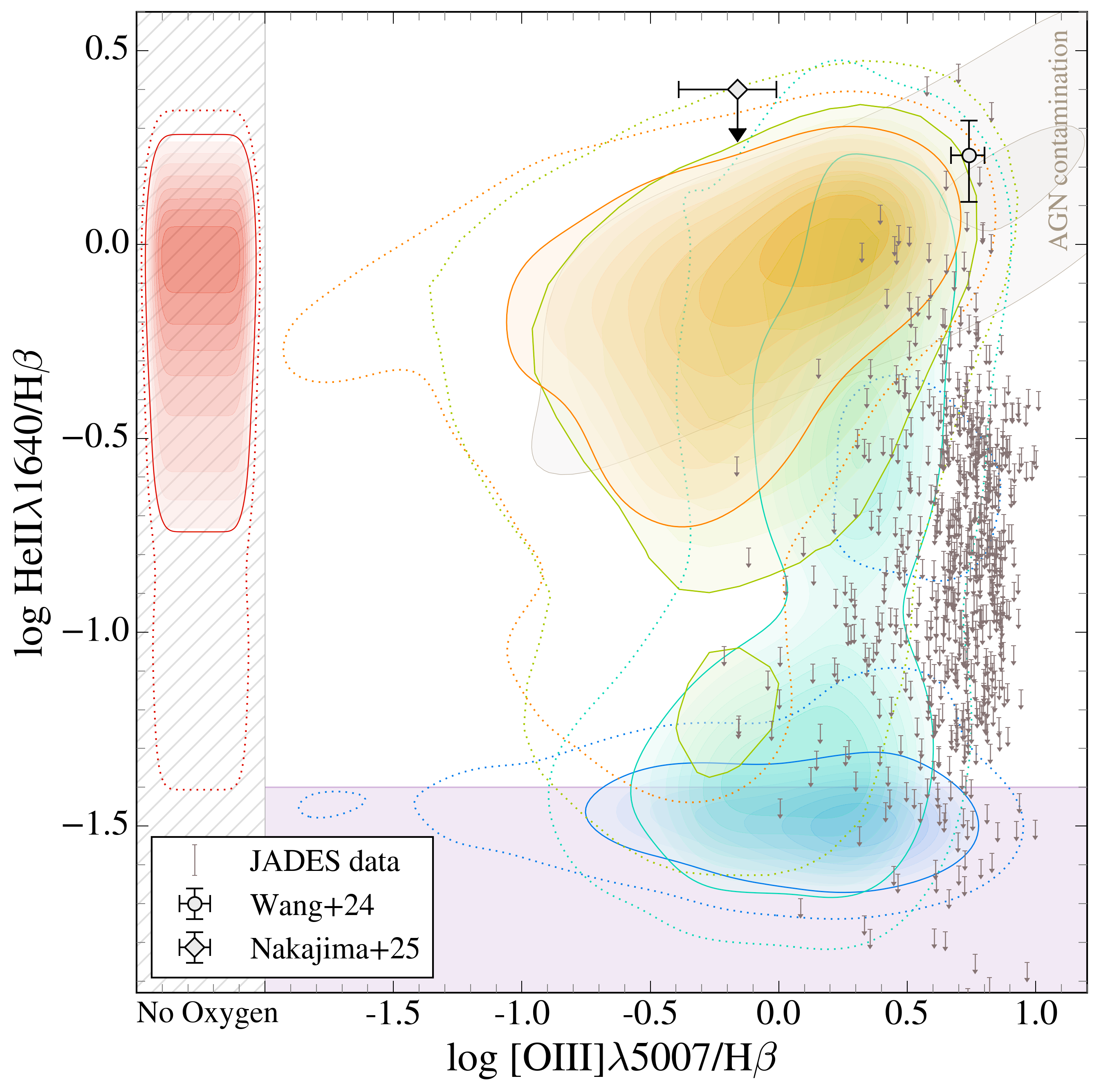} 
    \hspace{0.01\hsize}
    \caption{Same as Fig.~\ref{fig:fig1}, but for $\rm HeII\lambda1640 / H\beta$ versus $\rm [OIII]\lambda 5007/H\beta$. The grey shaded area represents the region containing $68\%$ of the AGN models from \citealt{feltre+16}. Points with errorbars are observations of LAP1-B \citep{Nakajima+25}, and RX J2129–z8HeII \citep{wang+24}. The grey arrows are $5 \sigma$ upper limits from JADES data \citep{Deugenio+25}.}
  \label{fig:fig2}
\end{figure}

Having determined the main phases of PopIII galaxies, we now study their synthetic emission constructed as described in Sec.~\ref{sec:2.2}. Given the rapid enrichment of PopIII galaxies, we choose emission line diagrams that combine characteristic signatures of metal-free and metal-enriched galaxies.
The most effective diagnostic of PopIII galaxies is the high equivalent width (EW) of HeII lines, especially the optical line at $\lambda = 4686 \rm \AA$ \citep[][dashed line in the right panel of Fig.~\ref{fig:fig1}]{Nakajima+22}.
The nebular oxygen line at $ \rm \lambda = 5007 \AA$ is instead particularly bright in typical star-forming galaxies, with the ratio $\rm [OIII]\lambda 5007 / H\beta$ (hereafter $\rm [OIII]/H\beta$) being used to determine the gas metallicity \citep[e.g.][]{maiolino+19}. 

On the right side of Fig.~\ref{fig:fig1} we show $ \rm EW(HeII\lambda 4686)$ versus $\rm [OIII]/ H\beta$, for $\rm logU = -1$ (see Appendix \ref{app:a} for other U). 
Our sample of pure PopIII galaxies with pristine gas has predominantly ($68\%$ of the total) high $\rm EW(HeII\lambda 4686) \geq 4 \AA$, and no oxygen lines. Instead, pure PopIII galaxies with self-polluted gas have the same characteristic HeII but simultaneously $0.1 <\rm [OIII]/H\beta < 4.5$.
This type of emission is maintained in the PopIII-rich phase and, partly, in the PopIII-mid. 
Instead, in the PopIII-poor phase, most galaxies have $ \rm EW(HeII) < 20 \AA$ and are thus populating the region strongly contaminated by AGN and black holes \citep{Nakajima+22}.
Ultimately, our results show that most galaxies with $>50\%$ of their $\rm M_\star$ in metal-free stars, including the pure PopIII ones, have the high $\rm EW(HeII) >4 \AA$ expected for pristine stars but also $\rm [OIII]/ H\beta \geq 0.1$, which can be associated with low-$Z$ PopII galaxies.

\begin{figure*}[thbp]
\centering
    \includegraphics[width=1.\hsize]{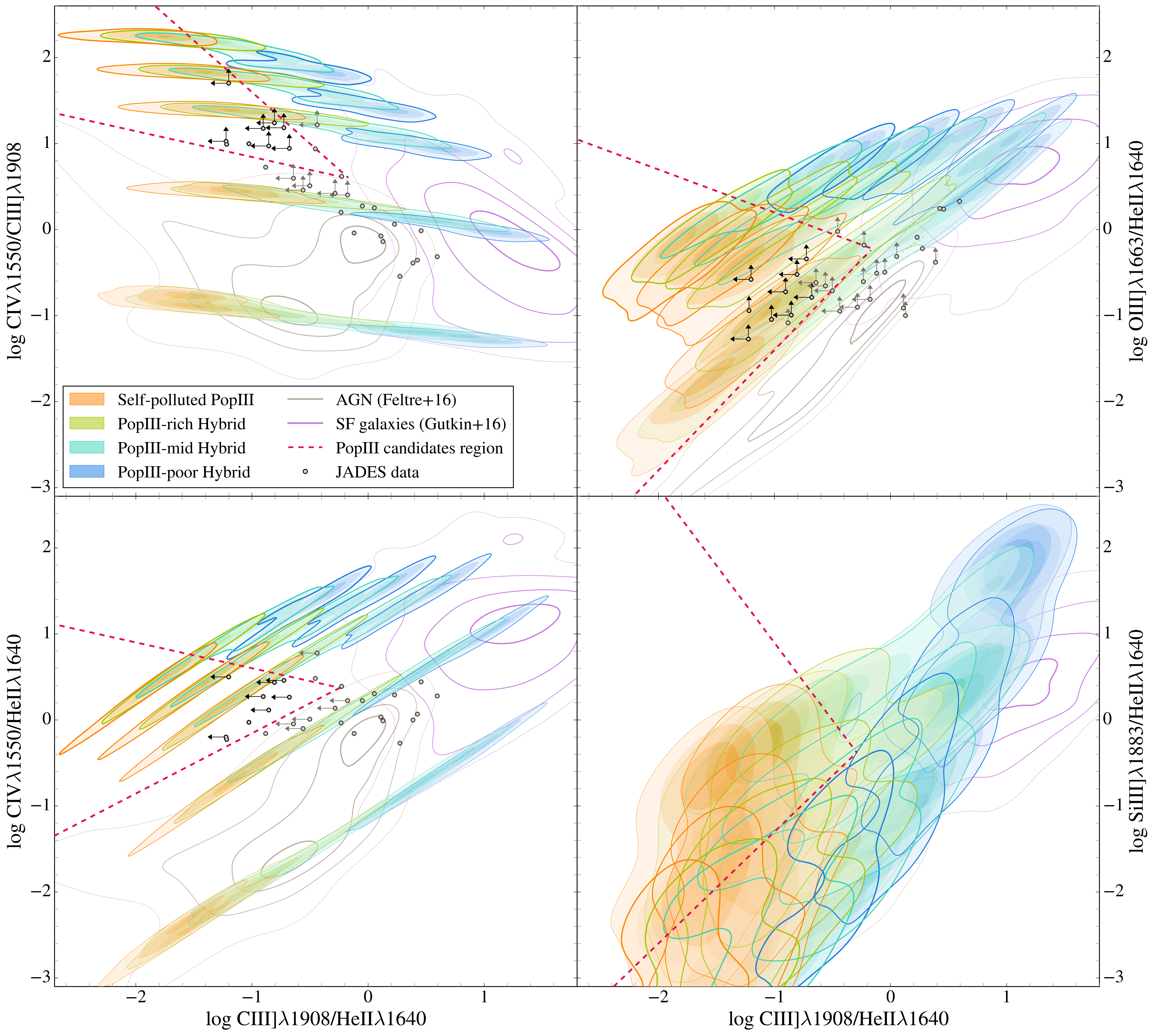} 
     \caption{Same as Fig.~\ref{fig:fig1}, but for different emission line ratios and $\rm logU=[-3,-2,-1,-0.5, 0]$. The thickness of the contours increases with the U value. The empty contours represent AGN models from \citealt{feltre+16} (grey) and star-forming galaxy models from \citealt{gutkin+16} (purple). The red dashed lines are our diagnostics for galaxies with $\rm >25 \% M_\star$ in PopIII stars (Sec.~\ref{sec:3.2}). The grey points are JADES data with tentative HeII and CIV detections ($S/N > 2$), with the darker shade highlighting the sources that we selected as possible PopIII candidates.}
  \label{fig:fig4}
\end{figure*}

In Fig.~\ref{fig:fig2} we show $\rm HeII\lambda1640 /H\beta$ versus $\rm [OIII]/H\beta$, for the same models of Fig.~\ref{fig:fig1}. Since the ultraviolet (UV) HeII line is the key observable used to claim the current PopIII candidates, we report the observations of LAP1-B \citep{Nakajima+25} and RX J2129–z8HeII \citep{wang+24}, which have measured $\rm H\beta$ fluxes. To visualize how AGN-like sources contaminate the diagram at high $\rm HeII/H\beta$ values, we also report the AGN photoionization models from \citealt{feltre+16}. 

The emission line ratios of LAP1-B are consistent with our PopIII galaxies, including those with $100\%$ metal-free stars. Still, given the upper limit on the HeII line, it remains compatible also with hybrid or PopII galaxies. 
Moving on to RX J2129–z8HeII, it can be reproduced by galaxies that have $\rm M_{\star, PopIII}> 25\% \rm M_\star$, with a higher likelihood for PopIII-rich galaxies. However, the observed $\rm [OIII]/H\beta$ is slightly higher than the bulk of our sample, lying in a region where the contamination from AGN is significant. 


In the same figure, we report the observed $\rm[OIII]/H\beta$ of galaxies from JADES \citep{eisenstein+23, Deugenio+25}, to showcase how several sources could populate the diagram regions of PopIII galaxies. All the $\rm HeII\lambda 1640/H\beta$ are $5\sigma$ upper limits; hence, these galaxies need further observations 
to determine whether they really are HeII emitters and to confirm their PopIII nature using the high EW of HeII (Fig.~\ref{fig:fig1}). Ultimately, our results show that high-$z$ sources should not be excluded as candidate PopIII systems solely based on high $\rm [OIII]$ emission, as has been done up to now. 

\subsection{PopIII diagnostics: identifying candidates}
\label{sec:3.2}

Here, we explore the distribution of PopIII galaxies in diagnostic diagrams involving UV metal emission lines observable at high-$z$, where we expect to find more PopIII-dominated systems. As outlined in Fig.~\ref{fig:fig1}, after the initial PopIII phase with pristine gas, PopIII SNe enrich the gas in the galaxies up to $Z_{gas} = [10^{-4.2} - 10^{-0.9}] Z_\odot$ and with a variety of different chemical patterns depending on the progenitor masses and SN explosion energies (e.g. $\rm -1 < [C/O] < 0.8$, $\rm -7 < [Si/O] < 0.8$).
Thus, we can use metal lines to pinpoint novel PopIII candidates. 
Hereafter, we will use the following notation: $\rm CIV\lambda 1550$ (figure) or simply $\rm CIV$ (text) for $\rm CIV\lambda \lambda 1548,1551 $, $\rm CIII]\lambda1908$ or $\rm CIII]$ for $\rm [CIII]\lambda1907 + CIII]\lambda1909$, $\rm OIII]\lambda 1663$ or $\rm OIII]$ for $\rm OIII]\lambda \lambda 1661,1666 $, $\rm SiIII]\lambda 1888$ or $\rm SiIII]$ for $\rm [SiIII]\lambda1883 + SiIII]\lambda1892$, and HeII for $\rm HeII\lambda1640$. 

\begin{figure}[t]
    \includegraphics[width=1.\hsize]{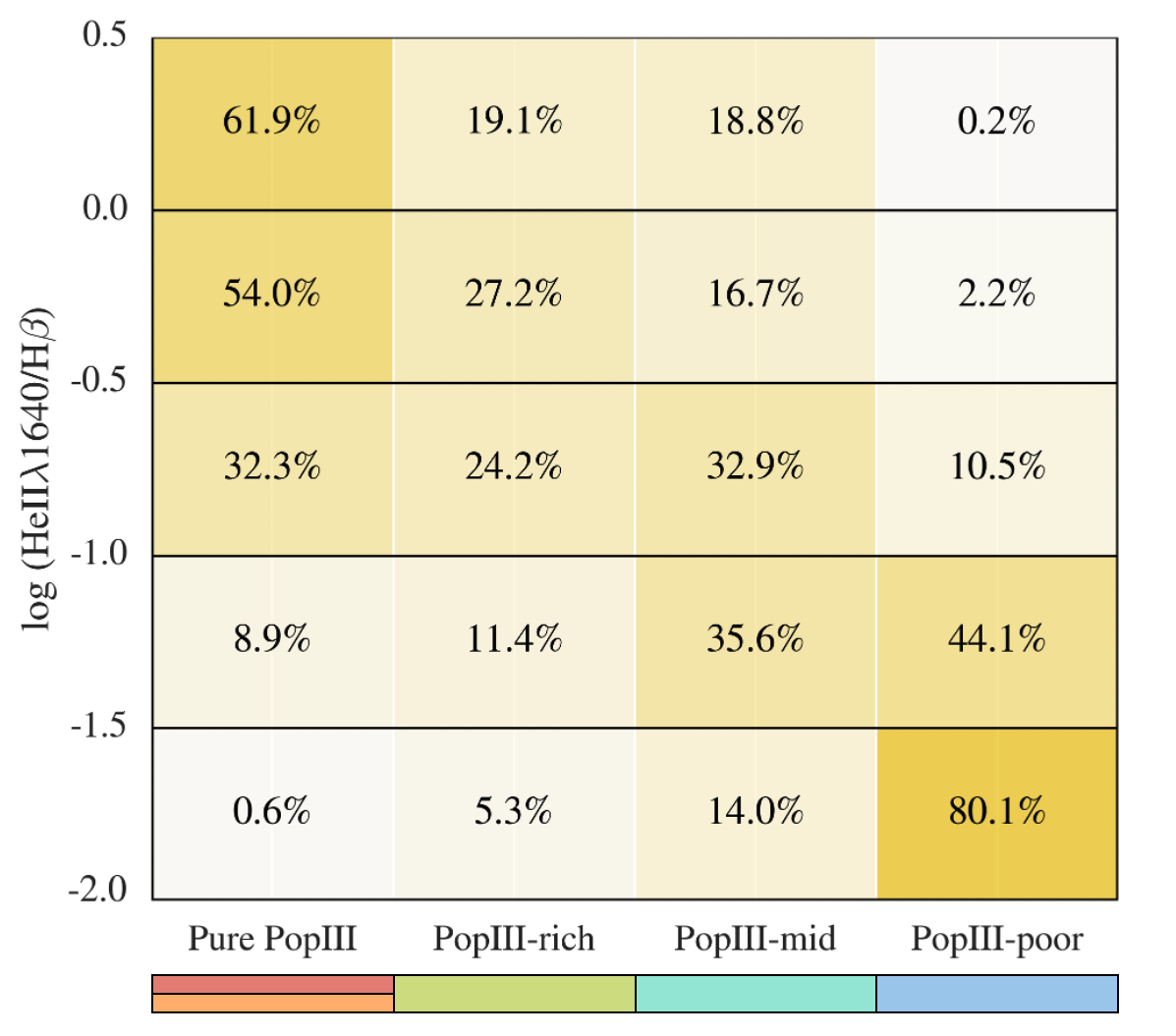} 
    \hspace{0.01\hsize}
    \caption{ Percentages of NEFERTITI PopIII galaxies in different evolutionary stages (see left side of Fig.~\ref{fig:fig1}) as a function of the $\rm HeII/H\beta$ emission. 
    }
  \label{fig:fig5}
\end{figure}

In Fig.~\ref{fig:fig4} we show how our PopIII galaxy models populate UV diagnostic diagrams involving different emission line ratios as a function of $\rm CIII]/HeII$, reporting for comparison models for AGN and normal PopII/I star-forming galaxies from \citealt{feltre+16} and \citealt{gutkin+16}. We see that pure self-polluted PopIII galaxies ($100\%$ of PopIII stars) are characterized by a low $\rm CIII]/HeII < 0.3$, which increases with the decreasing contribution of PopIII stars. 
However, for different U values, $\rm CIII]/HeII$ of pure PopIII galaxy models overlaps with that of PopIII-poor models ($<25\%$ of $M_{*}$ in PopIII stars),
creating degeneracies. Still, from the left-side diagrams of Fig.~\ref{fig:fig4} we can see that the CIV emission strongly depends on U, so its observed value can be used to determine the unknown ionization parameter (Appendix \ref{appendix:b} and see also \citealt{Nakajima+22}). Furthermore, if we restrict our analysis to models with $\rm logU > -2$, which are reasonable for highly-ionizing compact sources like PopIII galaxies, we can identify regions that are populated only by models with at least $25\%$ of $M_{*}$ in PopIII stars, without contamination from AGN or typical SF galaxies. Thus, PopIII galaxy candidates can be spotted by selecting objects that satisfy the following relations:

\begin{equation}
  \rm \frac{7}{2}\left(\frac{CIII]}{HeII} \right) ^{-0.3} < \frac{CIV}{CIII]} <\frac{5}{2}\left(\frac{CIII]}{HeII}\right)^{-1.2}
\end{equation}
\begin{equation}
  \rm \left( \frac{CIII]}{HeII} \right) ^{1.4}  < \frac{OIII]}{HeII} <\frac{1}{2} \left( \frac{CIII]}{HeII}\right)^{-0.5}
\end{equation}
\begin{equation}
  \rm \frac{7}{2}\left( \frac{CIII]}{HeII} \right) ^{0.7} < \frac{CIV}{HeII} <2 \left( \frac{CIII]}{HeII}\right)^{-0.3}
\end{equation}
\begin{equation}
  \rm  \left( \frac{CIII]}{HeII} \right) ^{1.3} < \frac{SiIII]}{HeII} <\frac{1}{8} \left( \frac{CIII]}{HeII}\right)^{-1.8}
\end{equation}

For reference, for $\rm log U > -2$, CIV and OIII] have $\rm EW \sim \rm 100 \AA$ while CIII] and SiIII] are much less intense, with $\rm EW \sim 1 \AA$ (see Appendix \ref{appendix:c}).

In Fig.~\ref{fig:fig4}, we also include galaxies from the JADES DR3 line flux catalog \citep{Deugenio+25}. To identify possible candidates, we select JADES galaxies with tentative detections of HeII and CIV, lowering the threshold at $S/N >2$. 
We find that 9 of the observed galaxies
dwell within our novel diagnostics exploiting metal-line ratios for self-polluted PopIII galaxies. These PopIII candidates are also those with the highest $\rm HeII/H\beta$ in Fig.~\ref{fig:fig2}. In particular, one of them has been previously identified as an AGN in \citealt{scholtz+25}, based on its UV high ionization lines. However, since this feature is consistent with self-polluted PopIII galaxies, which can have up to $\rm [NeIV] \lambda 2424 / CIII] \lambda 1909 \approx 1 $, we still include this object in our PopIII candidate sample.

\section{Discussion and Conclusions}
\label{sec: conclusions}

The quest for PopIII galaxies, now finally achievable with JWST, has yielded a few promising yet unconfirmed candidates. This scarcity of detections primarily stems from the extremely limited timeframe in which we can observe the emission from metal-free environments. 

In this Letter, we aimed to increase the number of PopIII candidates by accurately modeling the emission of early galaxies hosting first-generation stars.
To this end, we employed {\tt NEFERTITI} \citep{Koutsouridou2023}, a state-of-the-art model for simulating PopIII galaxies, which self-consistently follows the chemical enrichment of their initially pristine gas. 
The strength of this model is that it is calibrated with numerous local MW observations, including stellar archeology data, but it can also reproduce the properties of MW-like progenitors observed at high-$z$ \citep{rusta+24}. 

Our key finding is that pure PopIII galaxies can have metal emission lines due to their self-polluted gas, thus chemically enriched by the first PopIII SNe. Indeed, we find that galaxies with only metal-free stars have $\rm [OIII]/H\beta \approx 1$ during their first episodes of metal enrichment, before the onset of PopII formation. This crucial phase can be extended if we consider a temporal delay between PopIII SNe and the subsequent star formation, as implemented in other works \citep[e.g.][]{katz+23}. Our finding enlarges the pool of PopIII candidates to include more easily detectable sources with metal emission lines.

Secondly, we predict that PopIII galaxies can maintain their characteristic high HeII emission for up to $\rm \approx 20 \, Myrs$, hence not only in their initial pure PopIII stage
but also in the following hybrid phase, when PopIII and PopII stars co-exist. This is shown in Fig.~\ref{fig:fig5}, illustrating 
the results obtained using our {\tt NEFERTITI} sample of $\approx 1130$ PopIII galaxies. If $\rm HeII/H\beta> 0.1$, there is $ \gtrsim 90\%$ probability that the PopIII galaxy has at least $  \rm 25\% \, M_{\star, PopIIII}$. Instead, even at high $\rm HeII/H\beta> 1$, we can only say that it is a pure PopIII galaxy with $62\%$ probability, but there is still a fair chance that it can be a PopIII-dominated hybrid.


Next, we find that the PopIII candidates LAP1-B \citep{vanzella+23, Nakajima+25} and RX J2129–z8HeII \citep{wang+24} have $\rm HeII/H\beta$ and $\rm [OIII]/H\beta$ consistent with our models for PopIII galaxies, including pure self-polluted ones. However, based on these emission line ratios only, we cannot rule out AGN contamination, especially for RX J2129–z8HeII.



Finally, we propose novel diagnostics for identifying PopIII candidates at high-$z$, exploiting UV metal emission lines from PopIII-enriched gas in galaxies that host metal-free stars. 
We present simple relations between emission line ratios that can help us select pure PopIII and hybrid galaxies with $\rm M_{\star, PopIII}> 25 \% M_{\star}$, without contamination from AGN or PopII/I SF galaxies. However, to discriminate between PopIII stars or intermediate-mass black holes accreting pristine gas, a high $ \rm EW(HeII)$ is required \citep{Nakajima+22}. 

Among JADES data, we identified 9 systems that lie within our new diagnostics and for which we require deep observations 
to confirm their high HeII emission. 
The number of PopIII candidates found with our proposed method is quite remarkable, considering that we have only used one of the available JWST surveys for high-$z$ galaxies. 
Thus, this study brings us closer to the long-awaited detection of PopIII galaxies: now we need to observe our candidate sources with the outstanding sensitivity of JWST to confirm their primordial nature.

\begin{acknowledgments}
We thank the anonymous referee for their useful and positive comments. 
This project received funding from the ERC Starting Grant
NEFERTITI H2020/804240 (PI: Salvadori).
AM acknowledges support from PRIN-MUR project “PROMETEUS” financed by the European Union - Next Generation EU, Mission 4 Component 1 CUP B53D23004750006, and INAF funding through
the “Ricerca Fondamentale 2023” program (mini-grant 1.05.23.04.01). 

\end{acknowledgments}

\appendix

\section{Ionization parameter dependence}
\label{app:a}

\begin{figure*}[htbp]
    \includegraphics[width=1.\hsize]{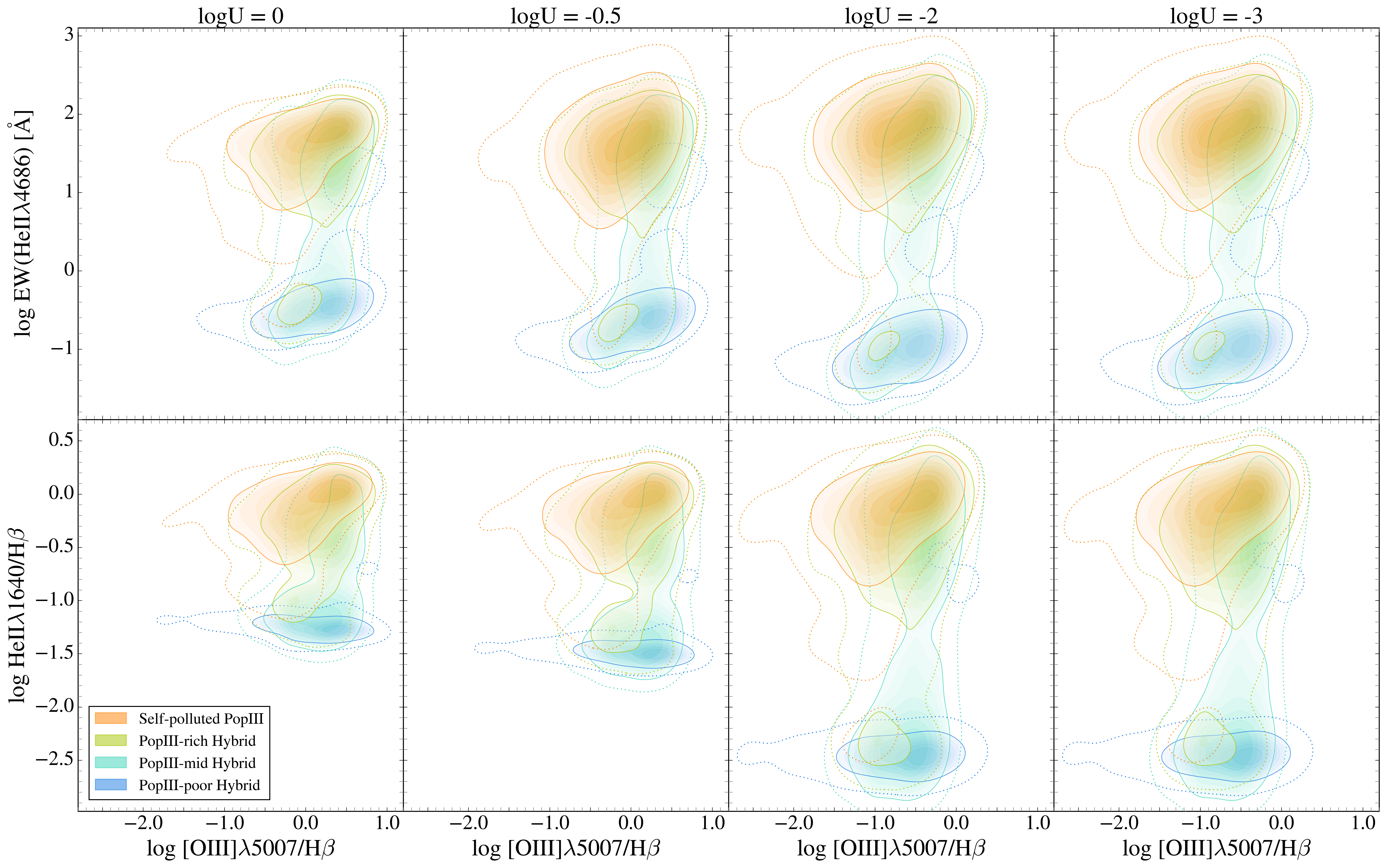} 
    \caption{Same as Fig.~\ref{fig:fig1} (top row) and Fig.~\ref{fig:fig2} (bottom row), but for different U values.}
  \label{fig:fig_a}
\end{figure*}

In Fig.~\ref{fig:fig_a} we report the distribution of {\tt NEFERTITI} PopIII galaxy models for different U values, in the $ \rm EW(HeII\lambda4686)$ versus $\rm [OIII]/H\beta$ diagram (top row) and in the $ \rm HeII/H\beta$ versus $\rm [OIII]/H\beta$ diagram (bottom row). We notice that the peaks of the distributions of $\rm EW(HeII\lambda4686)$ and $\rm HeII/H\beta$ remain approximately constant for different U. Instead, the $\rm [OIII]/H\beta$ decreases with decreasing U, reducing the likelihood of detecting PopIII galaxies with chemically-enriched gas. 

\section{Relations for the ionization parameter}
\label{appendix:b}

Here, we report relations between emission line ratios that can be used as proxies for the unknown U value. Indeed, on the left side of Fig.~\ref{fig:fig4}, we can identify linear relations for {\tt NEFERTITI} PopIII galaxy models with fixed U, in logarithmic scale:
\begin{equation}
  \rm  log(CIV/CIII]) = a_1(U) \cdot log(CIII]/HeII) + b(U)
\end{equation}
\begin{equation}
   \rm log(CIV/HeII) = a_2(U) \cdot log(CIII]/HeII) + b(U),
\end{equation}
where the dependence on U can be parametrized as follows:
\begin{equation}
    \rm a_1(U)=-0.02 \cdot logU -0.24
\end{equation}
\begin{equation}
    \rm a_2(U) = -0.02 \cdot logU +0.76
\end{equation}
\begin{equation}
    \rm b(U) = -0.16 \cdot (logU)^2 + 0.47 \cdot logU + 1.76.
\end{equation}

\section{Equivalent widths of UV metal lines}
\label{appendix:c}

\begin{figure*}[thpb]
    \includegraphics[width=1.\hsize]{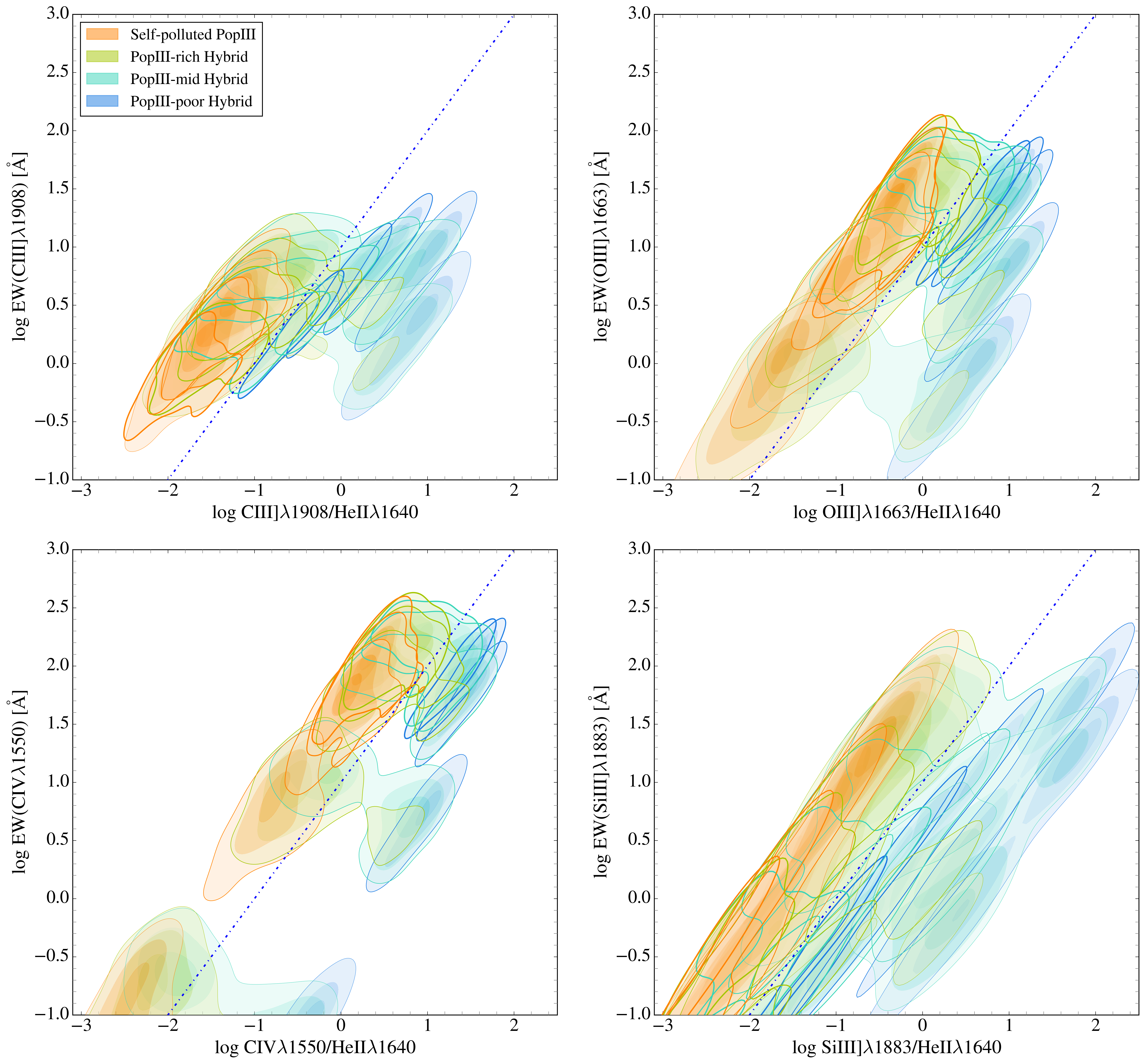}
    \caption{Same as Fig.~\ref{fig:fig4}, but for the EWs of UV metal lines versus their ratio with HeII. The black dotted lines identify regions populated only by models with $>25\%$ PopIII $\rm M_\star$.}
  \label{fig:appendix_c}
\end{figure*}

In Fig.~\ref{fig:appendix_c}, we report the EWs of CIII], CIV, OIII], and SiIII] versus their ratio with HeII. With increasing U, we notice that the EWs of CIV and OIII] are higher while those of CIII] and SiIII] become lower. Except for CIII], the other lines' EWs depend strongly on U, ranging over four orders of magnitude. For $\rm logU > -2$, our PopIII galaxy models have $\rm EW(CIII]) < 10 \AA $, $\rm EW(CIV) > 10 \AA$, $\rm EW(OIII]) > 1 \AA$ and $\rm EW(SiIII]) < 30 \AA$. Moreover, the following condition determines a region populated only by PopIII galaxies with $\rm M_{\star, PopIII}> 25 \% M_\star$: 
\begin{equation}
    \rm EW(X) > 10 \cdot X/HeII
\end{equation}
where X represents CIV, CIII], OIII], or SiIII]. 

\bibliography{refer, codes}
\bibliographystyle{aasjournalv7}

\end{document}